\documentclass{emulateapj}
\usepackage{apjfonts}

\newcommand{\psr}{J0737$-$3039}
\newcommand{\psrs}{PSRs~J0737$-$3039}

\newcommand{\msun}{\ifmmode M_{\odot}\else$M_{\odot}$\fi}
\newcommand{\rsun}{\ifmmode R_{\odot}\else$R_{\odot}$\fi}
\newcommand{\degrees}{\ifmmode^{\circ}\else$^{\circ}$\fi}
\newcommand{\amin}{\ifmmode^{\prime}\else$^{\prime}$\fi}
\newcommand{\asec}{\ifmmode^{\prime\prime}\else$^{\prime\prime}$\fi}
\newcommand{\kms}{\,{\rm km\,s^{-1}}}
\newcommand{\td}{\ifmmode{\Delta t_d}\else$\Delta t_d$\fi}
\newcommand{\nud}{\ifmmode{\Delta \nu_d}\else$\Delta \nu_d$\fi}
\newcommand{\bsh}{\mathbf{H}}
\newcommand{\bsv}{\mathbf{V}}
\newcommand{\bsr}{\mathbf{R}} 
\newcommand{\ma}{M_{\rm A}}
\newcommand{\mb}{M_{\rm B}}
\newcommand{\viss}{\ifmmode V_{\rm ISS}\else$V_{\rm ISS}$\fi}
\newcommand{\vplane}{\ifmmode V_{\rm plane}\else$V_{\rm plane}$\fi}
\newcommand{\vperp}{\ifmmode V_{\rm perp}\else$V_{\rm perp}$\fi}

\slugcomment{To be submitted to ApJ Letters}

\shorttitle{Systemic Velocity of PSRs J0737$-$3039}
\shortauthors{Ransom et al.}

\begin{document}

\title{Green Bank Telescope Measurement of the Systemic Velocity of
  the Double Pulsar Binary J0737$-$3039 and Implications for its
  Formation}

\author{
S.~M.~Ransom\altaffilmark{1,2},
V.~M.~Kaspi\altaffilmark{1,2,3},
R.~Ramachandran\altaffilmark{4},
P.~Demorest\altaffilmark{4},
D.~C.~Backer\altaffilmark{4},
E.~D.~Pfahl\altaffilmark{5},
F.~D.~Ghigo\altaffilmark{6},
D.~L.~Kaplan\altaffilmark{7}}


\altaffiltext{1}{Department of Physics, Rutherford Physics Building,
  McGill University, 3600 University Street, Montreal, Quebec, H3A
  2T8, Canada; ransom@physics.mcgill.ca}
\altaffiltext{2}{Department of Physics and Center for Space Research,
  Massachusetts Institute of Technology, Cambridge, MA 02139}
\altaffiltext{3}{Canada Research Chair; NSERC Steacie Fellow}
\altaffiltext{4}{Department of Astronomy, University of California,
  601 Campbell Hall, Berkeley, CA 94720}
\altaffiltext{5}{Chandra Fellow; Harvard-Smithsonian Center for Astrophysics, 60
  Garden Street, Cambridge, MA 02138}
\altaffiltext{6}{National Radio Astronomy Observatory, P.O. Box 2,
  Green Bank, WV 24944}
\altaffiltext{7}{Department of Astronomy, California Institute of
  Technology, MS 105-24, Pasadena, CA 91125}

\begin{abstract}
  We report on the measurement at 820- and 1400-MHz of orbital
  modulation of the diffractive scintillation timescale from pulsar A
  in the double-pulsar system \psr\ using the Green Bank Telescope.
  Fits to this modulation determine the systemic velocity in the plane
  of the sky to be \viss\ $\simeq$140.9$\pm$6.2$\kms$.  The parallel
  and perpendicular components of this velocity with respect to the
  line of nodes of the pulsar's orbit are \vplane\ 
  $\simeq$96.0$\pm$3.7$\kms$ and \vperp\ $\simeq$103.1$\pm$7.7$\kms$
  respectively.  The large \vperp\ implies that pulsar B was born with
  a kick speed of $\ga$100$\kms$.  Future VLBA determination of the
  angular proper motion in conjunction with improved \viss\ 
  measurements should provide a precise distance to the system.  Using
  high-precision timing data and the \viss\ model, we estimate a
  best-fit orbital inclination of $i=88\fdg7\pm0\fdg9$.
\end{abstract}

\keywords{binaries: general --- ISM: general --- pulsars: general ---
  pulsars: individual~(\objectname{PSR~J0737$-$3039}) --- stars:
  kinematics}

\section{Introduction}

For over twenty years observers have known that measurements of the
decorrelation bandwidth \nud\ and scintillation timescale \td\ of
pulsars undergoing strong Diffractive Interstellar Scintillation
(DISS) can be used to estimate their velocity in the plane of the sky
\citep{ls82}.  \citet{cr98} examined DISS-derived pulsar velocities in
detail and found that the measurements depend heavily on the observing
frequency, the direction and distance to the pulsar, and the detailed
distribution of the interstellar material causing the scintillation.
This last point makes measurements of pulsar velocities particularly
difficult since different models for the electron distribution and its
irregularities can cause differences in the ``measured'' scintillation
velocities (\viss) by factors of a few.  However, for binary pulsars
in compact orbits (orbital periods $P_{\rm orb} \la 1$\,day), the
pulsar-timing-derived orbital velocities can be used to calibrate
\viss\ measurements and remove many model-dependent and/or systematic
effects.  Unfortunately, suitable binary pulsars are rare and
successful measurements of this kind have only been made for two
pulsars: PSR~B0655$+$64 by \citet{lyn84} and PSR~J1141$-$6545 by
\citet*[hereafter OBvS]{obvs02}.

\psrs A \& B \citep[hereafter A and B;][]{bdp+03,lbk+04} comprise a
fantastic double-pulsar binary (hereafter 0737) consisting of the
22.7-ms pulsar A and the 2.77-s pulsar B.  It is nearby
($\sim$0.6\,kpc), mildly eccentric ($e\sim0.088$), compact ($P_{\rm
  orb}\sim2.45$\,h), highly inclined ($i\sim87\degr$), strongly
relativistic, and displays eclipses of A and very large but systematic
flux variations of B.  Its proximity, relatively high measured flux
density, and rapidly moving pulsars (orbital velocities
$\sim$300$\kms$), make it an ideal target for \viss\ studies.  In this
paper we report measurements of the orbital modulation of \viss\ from
A at 820- and 1400-MHz using data from the 100-m Green Bank Telescope
(GBT).

\section{Observations and Data Reduction}
\label{sec:obs}

In 2003 December, our group was awarded 5$\times$6\,hr Exploratory
Time tracks on 0737 with the GBT as part of the NRAO Rapid Science
program.  Three of the observations discussed in this paper were made
with the Berkeley-Caltech Pulsar Machine \citep[BCPM; e.g.][]{krb+04}
using summed IFs and 72\,$\mu$s sampling; one at 1400\,MHz using
96$\times$1\,MHz channels and the other two at 820\,MHz using
96$\times$0.5\,MHz channels (Table~\ref{tbl}).  During one 820\,MHz
observation, we also obtained data with the GBT Spectrometer SPIGOT
Card, a new, flexible, and high performance pulsar backend developed
at Caltech and NRAO. It processes auto-correlations from the GBT
Digital Spectrometer using two custom digital logic cards.  These
cards accumulate, sort, pack, and decimate the data before sending it
to a PC for further packing and output into FITS files on a multi-TB
RAID array.  The SPIGOT currently handles one or two IFs with
bandwidths of 50- or 800-MHz.  We used a 50\,MHz mode (\#44) to
process 1024\,lags from summed IFs that were 8-bit sampled every
40.96\,$\mu$s.  In total, the SPIGOT observation generated 420\,GB of
data which we recorded onto removable hard disks.

\begin{figure}
  \epsscale{1.15}
  \plotone{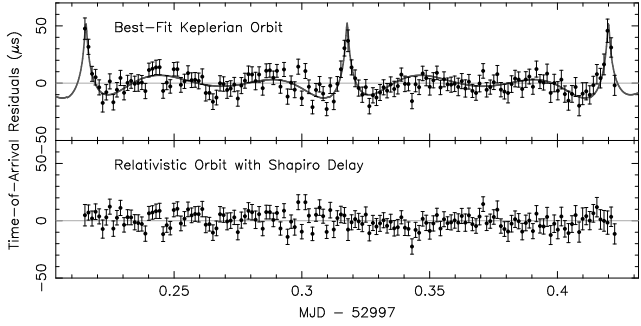}
\caption{
  Timing results from the 820\,MHz GBT$+$SPIGOT observation listed in
  Table~\ref{tbl}.  (Top) Residuals from a Keplerian orbit fit to the
  data.  Systematic deviations corresponding to the unmodeled Shapiro
  delay (shown as the line) are obvious in each orbit.  (Bottom)
  Residuals from a fit using the ``DD'' timing model which includes
  the Shapiro delay for a 1.25\,\msun\ pulsar B at an orbital
  inclination $i$=88\fdg4. The post-fit RMS residuals for the SPIGOT
  data alone are $<$7\,$\mu$s.\label{fig:shapiro}}
\end{figure}

We folded these data modulo the predicted pulse period given the
timing solution from \citet{lbk+04}.  These folds were used to
determine a local timing solution for A and to create dynamic spectra
from which to measure scintillation parameters.  For the timing
analysis, we measured 300 times-of-arrival (TOAs) from each of the
observations (for individual integration times of $\sim$1\,min) by
cross-correlating the folded profiles with a template profile from the
SPIGOT data and then referencing the maximum of the cross-correlation
with the time of the start of the observation as determined from the
observatory clock.  Our initial timing fits using {\sc
  TEMPO}\footnote{\url{http://pulsar.princeton.edu/tempo}} showed
approximately linear, frequency-dependent drifts in the TOAs of tens
of $\mu$s during each of the 5$-$6\,hr observations.  Since A's
pulsations are significantly linearly polarized \citep{drb+04}, such
apparent drifts may be caused by changes in the total-intensity pulse
profile with time due to the rotation of the receiver feed with
respect to the sky and gain differences in the two orthogonal IFs that
were summed.  In order to make high-precision measurements of {\em
  local} effects like the Shapiro delay, we fit for frequency jumps
between the observations which effectively whitened the residuals to
provide a post-fit RMS of $\sim$11\,$\mu$s.  Using a mass for B of
1.25\,\msun\ \citep[][]{lbk+04}, we determine the Shapiro ``shape''
parameter to be $\sin i = 0.99962^{+0.00038}_{-0.00095}$ which
corresponds to an inclination of 88\fdg4$^{+1.6}_{-1.4}$.  The errors
for $\sin i$ and the other orbital parameters were determined using a
bootstrap analysis \citep[e.g.][]{et86} which showed that our results
are consistent with those in \citet{lbk+04} at the 1-$\sigma$ level.
Figure~\ref{fig:shapiro} shows residuals from the 820\,MHz
GBT$+$SPIGOT observation alone.

\begin{deluxetable*}{ccccccccccccc}
  \tabletypesize{\scriptsize}
  \tablecaption{GBT Observations and \viss\ Model Fits for \psr A \label{tbl}}
  \tablewidth{0pt}
  \tablehead{\colhead{$f_{\rm ctr}$} & \colhead{Start} & \colhead{Freq.~Res.} & 
    \colhead{$T_{\rm obs}$} & \colhead{$T_{\rm int}$} & \colhead{$T_{\rm meas}$} & 
    \colhead{$\Delta\nu_{d}$} & \colhead{\viss} & \colhead{\vplane} & 
    \colhead{\vperp} & \colhead{$i$} &\colhead{} & \\
    \colhead{(MHz)} & \colhead{MJD} & \colhead{(bands$\times$MHz)} & 
    \colhead{(hr)} & \colhead{(s)} & \colhead{(s)} & \colhead{(MHz)} & 
    \colhead{(km/s)} & \colhead{(km/s)} & \colhead{(km/s)} & \colhead{(deg)} & 
    \colhead{$\kappa$} & \colhead{$\chi^2/{\rm DOF}$}}
  \startdata
  820\tablenotemark{a}  & 52997.214 & 1024$\times$0.04883 & 5.00 & 10.0 & 320 & 0.096(4) 
  & 139.0(4.5) & 95.1(3.8)  & 101.4(8.0) & 91.0(1.1) & 0.631(11) & 0.64 \\
  1400\tablenotemark{b} & 52984.195 & 96$\times$1.0 & 6.12 & 12.24 & 612 & 1.8(2) 
  & 163(16) & 107(13) & 123(27) & 92.9(3.5) & 0.693(48) & 1.50
  \enddata
  
  \tablecomments{$^a$\,SPIGOT data. $^b$\,BCPM data. All confidence
    intervals are 95\% statistical. The number of degrees of freedom
    for the model fits are ${\rm DOF}_{\rm 820}$ = $56-4$ = 52 and ${\rm
      DOF}_{\rm 1400}$ = $36-4$ = 32.  For these measurements, $i$ can
    range from 0$-$180\degr\ (see \S\ref{sec:obs}). The errors on the
    fitted parameters were determined from projections of the $\chi^2$
    space after adjusting the errors of the \viss\ measurements such
    that $\chi^2/{\rm DOF}$=1.}

\end{deluxetable*}

Assuming a uniform Kolmogorov scattering medium, \citet{cr98} derived
the velocity estimator
\begin{equation}
  \label{eqn:viss}
  \viss = 2.53\times10^4 \frac{\sqrt{D\ \nud}}{\nu\ \td}\, \mathrm{km\,s}^{-1},
\end{equation}
where $D$ is the distance to the source in kpc, \nud\ is the
decorrelation bandwidth in MHz, \td\ is the scintillation timescale in
seconds, and $\nu$ is the observing frequency in GHz.  Note that
unlike \td, which varies based on the relative velocities of the
Earth, scintillation screen, and pulsar, \nud\ is a property of the
screen itself and therefore does not vary significantly during an
observation.  The initial constant differs by factors of $\sim$1$-$3
depending on the structure function of the scattering medium/screen as
well as its location and extent.  The key point is that $\viss \propto
\td^{-1}$, with an overall scaling that can be {\em measured} by
fitting the known orbital velocity of the pulsar to the measured
changes in \td.

\begin{figure*}
  \epsscale{1.1}
  \plottwo{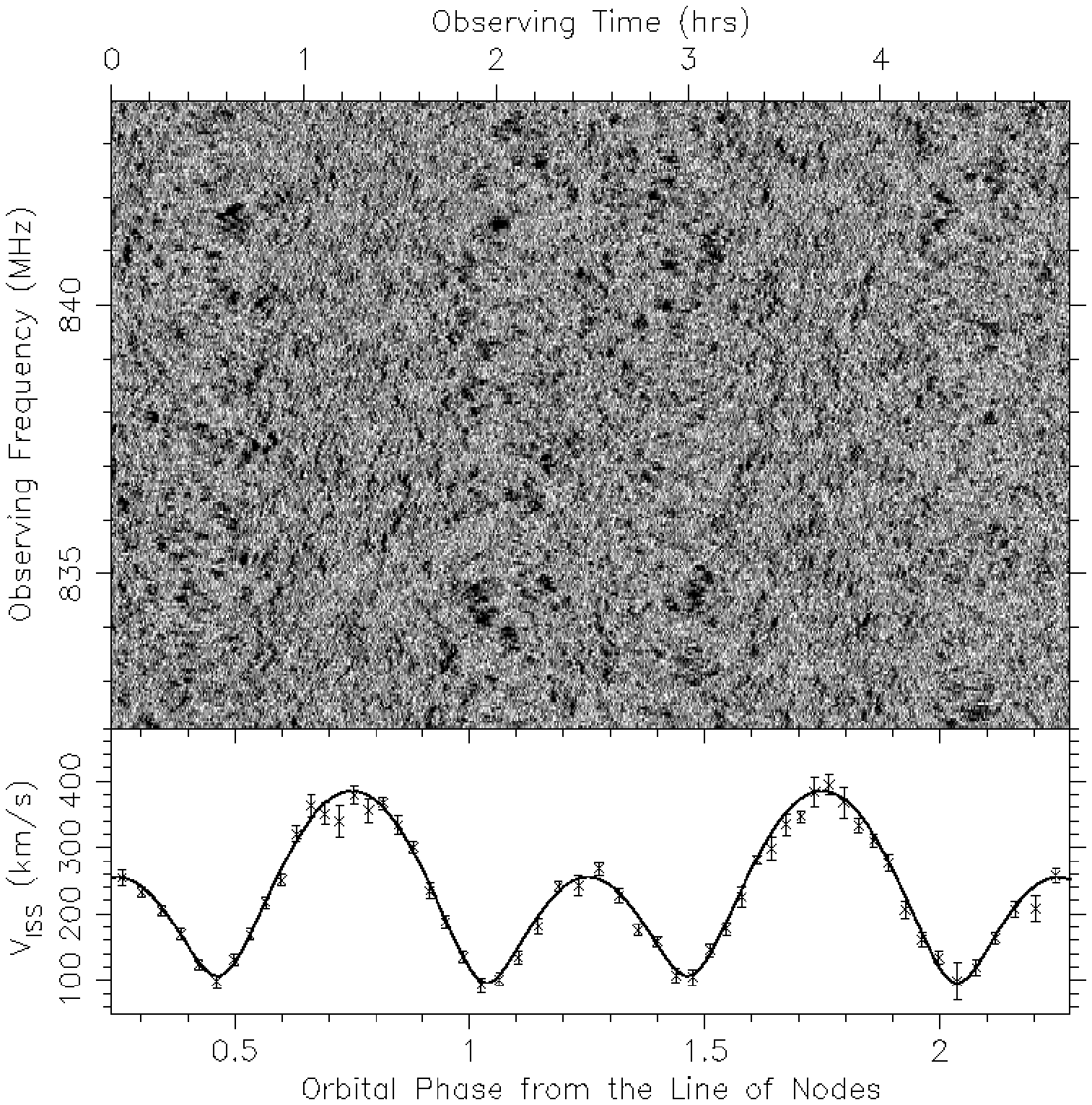}{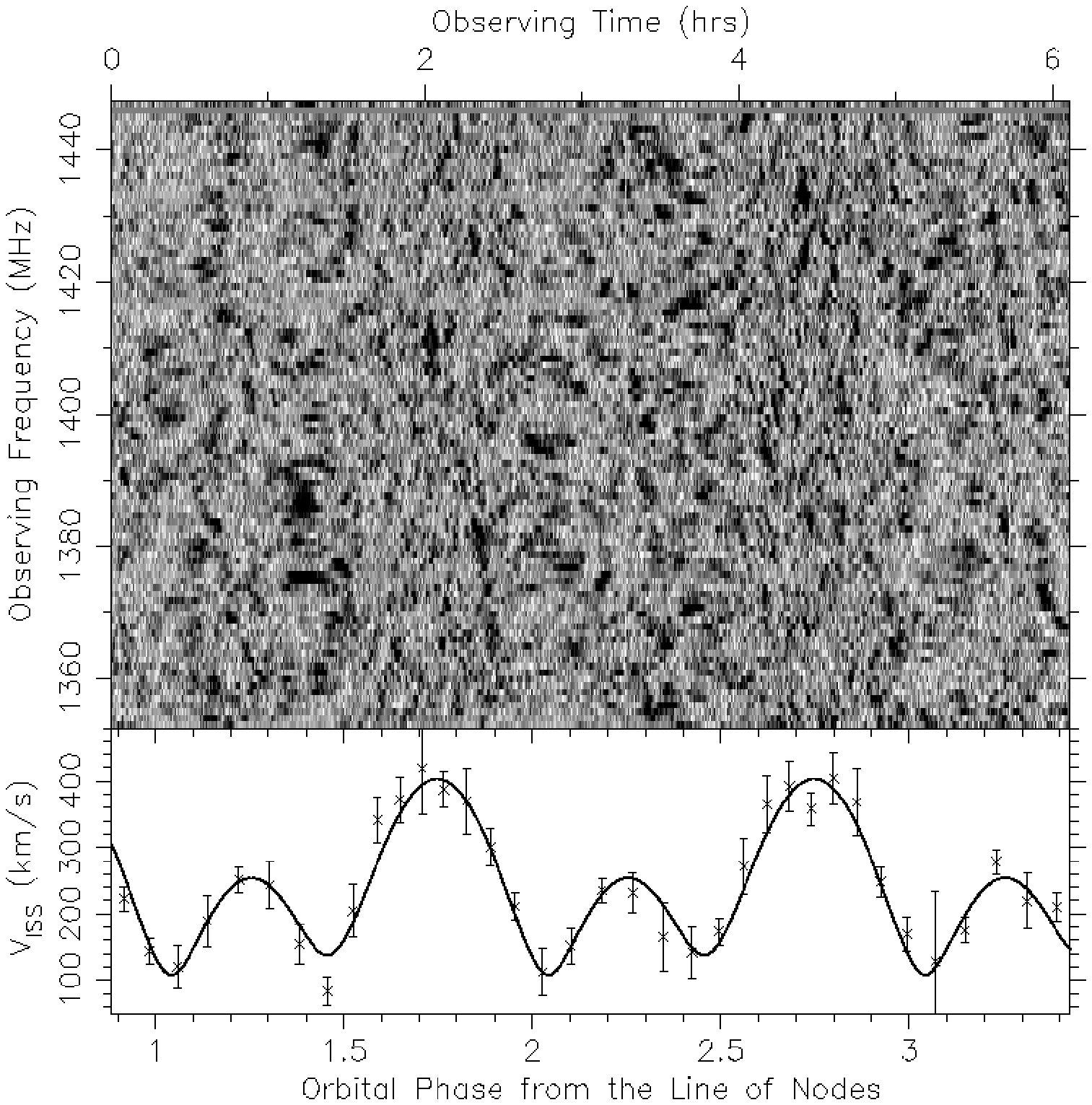}
\caption{
  Dynamic spectra (top) and \viss\ model fits (bottom) for the
  observations listed in Table~\ref{tbl}.  The 820\,MHz SPIGOT data is
  on the left and the 1400\,MHz BCPM data is on the right.  For the
  SPIGOT dynamic spectra, only the top 1/4 of the analyzed band is
  plotted.  Values $\pm$2-$\sigma$ from the average level were set to
  black/white respectively in order to enhance the contrast.
  \label{fig:viss}}
\end{figure*}

In order to determine \td\ and \nud\ for each of the observations
listed in Table~\ref{tbl}, we closely followed the process described
in OBvS.  Briefly, we created calibrated dynamic spectra (see
fig.~\ref{fig:viss}) by taking the on-pulse minus off-pulse flux
during a duration $T_{\rm int}$ divided by the average level in each
frequency channel during the full observation.  We then computed the
autocorrelation function (ACF) of non-overlapping (and hence
independent) blocks of the dynamic spectra of duration $T_{\rm meas}$ in
order to measure \td\ \citep[defined as the 1/$e$ half-width of the
central ACF peak in the time direction;][]{cor86}.  We measured \nud\ 
\citep[defined as the half-width at half-height of the central ACF
peak in the frequency direction;][]{cor86} and its errors by fitting
gaussians to the central ACF peaks from 15 sub-integrations and
examining their statistics.  We determined the relative errors for the
\td\ values similarly, by examining the statistics for \td\ fits made
in 6 and 3 frequency subbands for the SPIGOT and BCPM data
respectively.  Due to the small number of available subbands, we added
additional errors of 5 and 10\,km/s in quadrature to the measured
errors for the SPIGOT and BCPM data respectively to guard against
underestimates of \viss\ errors.

We converted the measured \td\ values into uncalibrated \viss\ 
measurements using eq.~\ref{eqn:viss}, and calculated the true anomaly
$\Theta$ at the center of each \viss\ measurement interval using the
timing ephemeris from \citet{lbk+04}.  We performed a weighted
least-squares fit of these data to eq.~7 in OBvS,
$\viss=\kappa(v^{\prime 2}+v^{\prime\prime 2})^{1/2}$, where
$v^{\prime}$ and $v^{\prime\prime}$ are the total velocities (i.e.
orbital + systemic) parallel to and perpendicular to the orbital line
of nodes in the plane of the sky respectively.  In contrast to OBvS,
we used the angle of periastron $\omega$ from the timing ephemeris and
fit for four parameters; the systemic velocities parallel to (\vplane)
and perpendicular to (\vperp) the line of nodes, the inclination $i$,
and the scaling parameter $\kappa$.  Fitting for $\kappa$ accounts for
uncertainties in the nature of the scattering medium and the distance
to the pulsar.  Table~\ref{tbl} lists the fitted parameters and their
95\% statistical confidence limits as determined from projections of
the $\chi^2$ space.  The knowledge that we observe 0737 nearly edge-on
($i\sim90\degr$) removes the degeneracy in possible inclination angles
seen by \citet{lyn84} and OBvS and produces a single solution for each
observation.  The errors in the fitted parameters were calculated
after scaling the \td\ errors such that the reduced-$\chi^2$ of the
fits equaled one.  Combining the BCPM and SPIGOT results provides our
best estimates of \vplane\ $\simeq$96.0$\pm$3.7$\kms$ and \vperp\ 
$\simeq$103.1$\pm$7.7$\kms$ for a total systemic velocity in the plane
of the sky of \viss\ $\simeq$140.9$\pm$6.2$\kms$.

We note that \viss\ fits can result in $i>90\degr$.  In that case (as
we have measured), the direction of \vperp\ is either aligned with the
orbital angular momentum vector if it is pointing slightly towards us,
or anti-aligned if the angular momentum is pointing slightly away from
us.  Combining the timing- and scintillation-based measurements gives
an inclination (in the more standard range 0\degr$-$90\degr) of
$i=88\fdg7\pm0\fdg9$.

\subsection{Possible Biases to the Measured \viss}

While the ``self-calibrating'' nature of binary pulsar \viss\ 
measurements eliminates many of the uncertainties found in \viss\ 
studies of isolated pulsars \citep[e.g.][]{cor86}, the motion of the
Earth with respect to the pulsar and the differential rotation of the
Galaxy both contribute to the measured \viss.

A flat, 220$\kms$, Galactic rotation curve implies a velocity
component in the plane of the sky of $\Delta V_{g\perp} \simeq
14(D/0.6\,{\rm kpc})\kms$ due to differential Galactic rotation in the
direction of 0737 ($l=245\fdg2$, $b=-4\fdg5$).  However, as
\citet{cr98} point out, the ISM rotates along with the Galaxy, and we
calculate that the {\em effective} transverse velocity is $<$1$\kms$
for all reasonable distances to 0737.

Earth's orbital motion projected onto the sky towards 0737 contributes
a doubly periodic velocity component $\Delta V_{\Earth\perp}$ to the
measured \viss.  For 0737, $\Delta V_{\Earth\perp}$ has minimum and
maximum values of 23 and 30$\kms$ respectively.  During these
observations, $\Delta V_{\Earth\perp} \sim 29\kms$.

Since we do not know the orientation of the measured \viss\ on the
plane of the sky, we cannot currently remove either of these biases
from our data.  However, measurement of an annual variation in \viss\ 
would allow us to remove $\Delta V_{\Earth\perp}$.  Alternatively,
VLBA determination of the proper motion direction will allow us to
subtract both biases directly.  In addition, the combination of the
VLBA angular proper motion $\mu$ with the transverse velocity
$V_\perp \simeq \viss$ will provide a unique geometric distance to the
pulsar ($D = V_\perp / \mu$).

\section{Discussion}

The relatively large $\viss\simeq140\kms$ measured for 0737 has
important implications for the formation of the system.  Here we
assume that 0737 followed the standard formation scenario for double
neutron stars (DNSs) in the Galactic disk \citep[e.g.][]{bv91}.  In
this picture, the first NS (A) forms in orbit about a massive
($\ga$8\,\msun) stellar companion.  The binary acquires some velocity
due to impulsive mass loss in the supernova explosion and a possible
natal kick to the NS.  At this stage, the system resembles a high-mass
X-ray binary (HMXB).  Detailed studies \citep[e.g.,][]{prps02} find
expected HMXB systemic speeds $\la$30$\kms$, because the NS kick is
distributed over the large total mass of the binary.  It is unlikely
that the first supernova provided the high systemic velocity implied
by the measured \viss.

At the end of the HMXB phase of DNS formation, the stellar companion
evolves, fills its Roche lobe, and transfers matter to the NS.
Because of the extreme mass ratio of the binary, the transfer is
dynamically unstable, and leads to a common-envelope phase.  If
coalescence is avoided, the NS emerges in a tight orbit with the
hydrogen-exhausted core of its companion.  The stellar core continues
its nuclear evolution, and may stably transfer matter to the NS via
Roche-lobe overflow. This is the favored mechanism for spinning up and
reducing the dipole magnetic field strength of A \citep{dv04}.
Subsequently, the stellar core explodes and produces the second NS
(B).  In 0737, the pre-supernova mass of the stellar core was likely
$M_c\simeq 2.0$$-$$3.5$\,$\msun$ \citep{dv04,wk04}.  The combination
of supernova mass loss and a kick to B could easily have given 0737 a
speed of $\ga$140$\kms$.  We now use the interestingly large component
(\vperp\ $\simeq$103$\kms$) of the systemic velocity perpendicular to
the orbital plane to constrain the magnitude of the kick imparted to
B.

The masses of A and B are $\ma \simeq 1.34$\,\msun\ and $\mb \simeq
1.25$\,\msun\ respectively, giving a total binary mass of $M_t = \ma +
\mb \simeq 2.59$\,\msun\ \citep{lbk+04}.  We assume that the
pre-supernova orbit was circular, and denote by $\bsr$, $\bsv$, and
$\bsh = \bsr\times\bsv$ the position, velocity, and orbital angular
momentum of the pre-collapse progenitor of B relative to A.  If B
receives an impulsive natal kick velocity of $\bsv_k$, the
post-supernova relative orbital velocity and angular momentum are
given, respectively, by $\bsv' = \bsv + \bsv_k$ and $\bsh' = \bsh +
\bsr\times\bsv_k$.  The velocity acquired by the binary center of mass
after the supernova is
\begin{equation}
\label{eq:vcm}
  \bsv_{\rm CM} = - \frac{\ma}{M_t}\frac{(M_c - \mb)}{(M_c + \ma)}\bsv
  + \frac{\mb}{M_t}\bsv_k ~.
\end{equation}
If $\bsv_k = 0$, then $\bsv_{\rm CM}$ is in the orbital plane.
Therefore, the large observed perpendicular component of \viss,
\vperp, may {\em require} B to have been born with a significant kick
velocity.  If we assume that $\bsv_{\rm CM}$, and thus the observed
\vperp, is entirely due to the second supernova explosion, it is
straightforward to show that \vperp\ is related to $\bsv_k$ by
\citep[see also][]{wkk00}
\begin{equation}
  \bsh'\cdot\bsv_{\rm CM} = H'\,\vperp  = \bsh\cdot\bsv_k 
  \left(1 + \frac{\ma}{M_c}\right)^{-1}~.
\end{equation}
For specified masses and orbital parameters before and after the
supernova, $V_k$ is minimized when $\bsv_k$ is parallel to $\bsh$
(i.e., perpendicular to the pre-supernova orbital plane).  The minimum
kick speed is then
\begin{equation}
  V_{k,{\rm min}} = \vperp\frac{H'}{H}
  \left(1 + \frac{\ma}{M_c}\right)~.
\end{equation}
Note that
\begin{equation}
  \frac{H'}{H} = \left[\left(\frac{M_t}{M_c + \ma}\right)
    \frac{a'}{a}(1-e'^2)\right]^{1/2}~,
\end{equation}
where $a'$ and $e'$ are the immediate post-supernova semimajor axis
and eccentricity, respectively, which have since evolved to their
observed values under the action of gravitational wave emission.
\citet{dv04} and \citet{wk04} each find that $e' \la 0.14$, so that
$a'/a$ cannot differ substantially from unity.  For our purposes, it
is sufficient to assume that $(a'/a)(1-e'^2)\simeq 1$.  Finally, we
obtain
\begin{equation}
  V_{k,{\rm min}} \simeq \vperp\left(\frac{M_t}{M_c}\right)^{1/2}
  \left(1 + \frac{\ma}{M_c}\right)^{1/2}~.
\end{equation}
For the range of pre-collapse core masses given above, and \vperp\ 
$\simeq$100$\kms$, we find that $V_{k,{\rm min}} \simeq
100$$-$150$\kms$, where the upper limit corresponds to $M_c = 2\msun$.
If the pre-supernova core mass takes its smallest possible value of
$M_c = \mb$, then $V_{k,{\rm min}} \simeq 213\kms$.

\citet{ps04} point out that at a distance of 0.6\,kpc, 0737 is only
$\sim$50\,pc from the Galactic plane.  They suggest that in order to
find 0737 so close to the plane, $\bsv_{\rm CM}$ must be very small
($\la$15$\kms$ or $\la$150$\kms$ at 68\% or 95\% confidence
respectively), and that the supernova mass loss and kick to B must
likewise have been very small.  The measurement of a large \viss\ 
implies that (1) either the velocity direction is largely in the
Galactic plane or we happen to see it as it passes through the plane,
and (2) moderate pre-collapse core masses and NS kicks are permitted
if not required.

VLBA observations will soon determine the magnitude and direction of
0737's proper motion $\mu\simeq0.05$\asec\,yr$^{-1}
(\viss/140\kms)(D/0.6\,{\rm kpc})^{-1}$.  For long-term timing
observations, the large velocity in the plane of the sky $V_\perp
\simeq \viss$ will cause an apparent acceleration \citep{shk70} that
will bias the measured spin- and orbital-period derivatives.  For
0737, $\dot P/P = V_\perp^2/(Dc) = V_\perp \mu/c \simeq
3.5\times10^{-18}$\,s$^{-1}$, which amounts to $\sim$5\% of A's
measured spindown and $\sim$2.5\% of the predicted orbital decay rate
due to gravitational wave emission.  Future bias-corrected
measurements of \viss\ should determine $V_\perp$ to a few percent and
VLBA observations will likely measure $\mu$ to even better precision.
Therefore, the ``Shklovskii effect'' should be measurable to $\la$5\%
precision and consequently, its error will contaminate the measurement
of the orbital period-derivative $\dot P_{\rm orb}$ due to the
emission of gravitational wave emission by $\la$0.2\%.

\section{conclusions}

We have measured the systemic velocity and inclination of 0737 using
the ``self-calibrating'' method of binary scintillation velocity
measurements.  The inferred high velocity \viss$\simeq$140$\kms$
strongly suggests that a substantial ($\ga$100$\kms$) kick was
imparted to pulsar B at its birth.  The large \viss\ will impact
long-term timing of the system and will allow a precise geometric
distance measurement when combined with a VLBA proper motion.

Future scintillation observations using the high frequency resolution
and wide bandwidth of the GBT$+$SPIGOT will allow substantially
improved measurements of \viss\ at 1400 and even 2200\,MHz where we
have already detected the orbital modulation of A in BCPM data.  These
measurements should also allow the removal of the contaminating
effects of the Earth's motion and the differential Galactic rotation.
Finally, we have already detected scintillation from B during the
bright portions of its orbit, but future observations may allow
``snapshot'' calibrations of \viss\ based on measurements of \td\ from
A and B at an instant in time given our knowledge of the relative
orbital velocities of the pulsars.  Such measurements would demand far
less telescope time.

{\em Acknowledgements} We extend special thanks to Karen O'Neil, Rich
Lacasse, Glen Langston, and Chris Clark for help with the
observations.  The National Radio Astronomy Observatory is a facility
of the National Science Foundation operated under cooperative
agreement by Associated Universities, Inc.  V.M.K. acknowledges
support from NSERC Discovery Grant 228738-03, NSERC Steacie Supplement
268264-03, a Canada Foundation for Innovation New Opportunities Grant,
FQRNT Team and Centre Grants, and NASA Long-Term Space Astrophysics
Grant NAG5-8063.

\bibliographystyle{apj}
\bibliography{apj-jour,pulsars,psrrefs,0737}

\end{document}